# Enhanced electron yield from a laser-plasma accelerator using high-Z gas jet targets

Mohammad Mirzaie[1], Nasr A. M. Hafz[1,*], Song Li[1], Thomas Sokollik[1], Fei He[2], Ya Cheng[2],

Zhengming Sheng and Jie Zhang[1,*]

[1]*Key Laboratory for Laser Plasmas (Ministry of Education) and Department of Physics and Astronomy, Shanghai Jiao Tong University, Shanghai 200240, China*

[2]*State Key Laboratory of High Field Laser Physics, Shanghai Institute of Optics and Fine Mechanics, Chinese Academy of Sciences, Shanghai 201800, China*

[*]*Corresponding authors: nasr@sjtu.edu.cn and jzhang1@sjtu.edu.cn*

An investigation of the multi-hundred MeV electron beam yield (charge) form helium, nitrogen, neon and argon gas jet plasmas in a laser-plasma wakefield acceleration experiment was carried out. The charge measurement has been made via imaging the electron beam intensity profile on a fluorescent screen into a 14-bit charge coupled device (CCD) which was cross-calibrated with nondestructive electronics-based method. Within given laser and plasma parameters, we found that laser-driven low Z- gas jet targets generate high-quality and well-collimated electron beams with reasonable yields at the level of 10-100 pC. On the other hand, filamentary electron beams which were observed from high-Z gas jets at higher densities reached much higher yield. Evidences for cluster formation were clearly observed in high-Z gases, especially in the argon gas jet target where we received the highest yield of ~ 3 nC. ©2014 Optical Society of America

## 1. INTRODUCTION

Laser-plasma wakefield acceleration (LWFA) was proposed in 1979 [1] and had a rapid and remarkable progress over the past decade, growing along with the advances in the technology of table-top ultra-short high power laser systems[2]. The concept of the LWFA can be described as follows: a tightly focused ultra-short terawatt laser pulse propagating through an under-dense plasma excites a large-amplitude (~100 GV/m) electrostatic relativistic plasma wave (with a phase velocity ~ c), thanks to the strong ponderomotive force of the laser pulse. This plasma wave which oscillates at the plasma frequency $\omega_p$ can accelerate electrons to relativistic energies within very short distances. Thus, the LWFA scheme has, in principle, the potential to be the basis for a new particle accelerator technology for producing ever compact and affordable electron accelerators. High quality electron beam acceleration by the LWFA mechanism depends on the laser-plasma parameters such as laser intensity, focal spot size, plasma density, interaction length and laser pulse duration. Generation of quasimonoenergetic electron beams with energies of the order of 100 MeV to GeV from laser-driven gas jets and discharged capillary was demonstrated[3-14]. In addition, applications to X-ray and gamma ray generation were proposed from the achieved LWFA electron beams[15-20]. From the applications point of view, higher electron beam charge is very important, so one has to find ways to enhance the electron beam charge from LWFA. It is also important to measure the charge by a reliable diagnostic technique. Many electron beam charge diagnostics are commonly used [3-4, 21-23], some of them were compared or cross calibrated with others[24-25]. Calibrated integrating current transformer (ICT) is widely used due to its simple set up and reliability and low cost[18, 26-28].

In this paper, we present experimental results focused on the electron beam charge/yield measurement from a LWFA employing various gas jets individually aiming simply at finding the effect of gas type on the beam charge. In the experiment we used a TW laser interacting with 4 mm-long supersonic jet of helium, nitrogen, neon, and argon gas jets individually. The generated electron beam charge was measured using electronics based on an integrating current transformer (ICT). The electron beam spatial profile was simultaneously recorded on a DRZ fluorescent screen which was imaged into a 14-bit CCD camera. The recorded images were processed using a MATLAB code and the ICT was cross-calibrated with the DRZ-CCD system. The experimental setup is described in Sec. II. and the image processing method is explained in Sec. III. Results and discussions are presented in Sec. IV and a conclusion is given in Sec. V.

## 2. EXPERIMENTAL SETUP

Our LWFA experiment was carried out at the newly-established Key Laboratory for Laser Plasmas (LLP) at Shanghai Jiao Tong University (SJTU), Shanghai, in China. We have a tabletop 200 TW, 10Hz, Ti: Sapphire 30 fs laser facility; however we haven't used the full power of the laser system in the present study. The experimental setup is shown in Fig. 1. Linearly polarized 800 nm laser pulses with energies of 0.5–1.5 J (~17–50 TW) were used. The 10.5 cm diameter laser beam was focused using 2 m focal length off-axis parabolic (OAP) mirror (TYDEX Research and Industrial Optics). The full-width at half maximum (FWHM) of the focused laser spot size was measured to be 33 µm, so the $1/e^2$ intensity radius of the focal spot was 28 µm giving a Rayleigh length $Z_r$ of 3 mm. The Strehl ratio of the focal spot was 0.4–0.5. The peak focused laser intensity and the corresponding normalized vector potential, $a_0$, were approximately 1.4–4.2×10$^{18}$ W.cm$^{-2}$ and 0.8–1.4, respectively. A 4-mm-long supersonic nozzle gas-jet target having a Mach number of 5 (Smart shot LX-03R, Smart Shell Co., Ltd.)[29-32] was used. The stagnation pressure of the gas jet was varied from 2–25 atm. At such gas pressure range and at the above-mentioned laser intensities, the plasma density ($n_e$) was estimated to be 3.35×10$^{18}$–4.0×10$^{19}$ cm$^{-3}$ depending on gas type. Ionization of helium, nitrogen, neon and argon atom into He$^{2+}$, N$^{5+}$, Ne$^{6+}$ and Ar$^{8+}$ based on ionization experiments[33] and the barrier-suppression model[34] are obtainable, respectively. To measure the accelerated electron beam charge, an integrating current transformer (ICT 0550-070, 20:1, Bergoz) was used. An aluminum foil with a thickness of 12.5 µm was placed between the gas jet and the ICT to stop laser beam after interaction. To trigger and synchronize the ICT with the laser pulse, a delay/pulse generator (DG535 Stanford Research System) and a photodiode were used. An oscilloscope (DSO-X-3024A Agilent Technologies) was used to monitor the ICT output signal. We used a DRZ fluorescent screen (DRZ-PLUS, Mitsubishi Chem. Corp.) and a 14-bit CCD camera (PCO Pixelfly qe usb) coupled with an objective lens for imaging the spatial profile of the accelerated electron beams. The energy spectra of the accelerated electron beams were measured by a single shot electron spectrograph composed of C-shaped dipole magnet. The magnet is composed of two rectangle permanent magnets of 6 (length) × 4 (width) cm$^2$ with 1 cm pole gap, to deflect the electron beam on the DRZ screen placed 17.3 cm away from the magnet end, the effective magnetic field strength was ~ Beff = 0.9 T. The field between the pole pieces has been measured and fed into a Matlab code which calculates the relativistic electron beam trajectory given the measured 2D field map. The electron beam pointing angle at the magnet entrance plane has been stabilized [35–38], and fed into the code thus the energy spectrum has been measured properly. For monitoring the laser-plasma interaction volume, we imaged the 2ω nonlinear Thomson scattered laser light into a 14-bit CCD camera.

## 3. ICT-DRZ CROSS-CALIBRATION

### 3.1. Charge measurement by the ICT

Charge measurement using the ICT is a popular method which is being used by many experimentalists as a nondestructive, low cost and fast charge measurement device. We used an ICT to measure the accelerated electron beam charge as our reference charge measurement instrument. The ICT was installed in vacuum at the distance of 80 cm from the gas jet. As long as the electron beam passes through the ICT, it generates an electric signal which will be sent to a beam charge monitoring unit (BCM) for processing. The electric signal generated by the ICT due to the passage of the electron beam should fall within either one of two integration windows of the BCM unit; the integration window (W1) of the timing view signal is shown in Fig. 2(d). The BCM unit needs to be triggered with delay from laser-plasma interaction time then it integrates the ICT signal and the final output will be generated as shown in Fig. 2(a). The voltage difference $\Delta V=5.95$ mV in Fig. 2(a) corresponds to a beam charge of 59.16 pC, given the BCM unit gain of 6 dB. To guarantee the right adjustment for the BCM trigger time, we used a fast photodiode to monitor the arrival of the laser pulse to the interaction region Fig 2. (b). We measured electron beam charge for 124 laser shots. In these measurements, the four gas jets of helium, nitrogen, neon, and argon were used as acceleration media, independently.

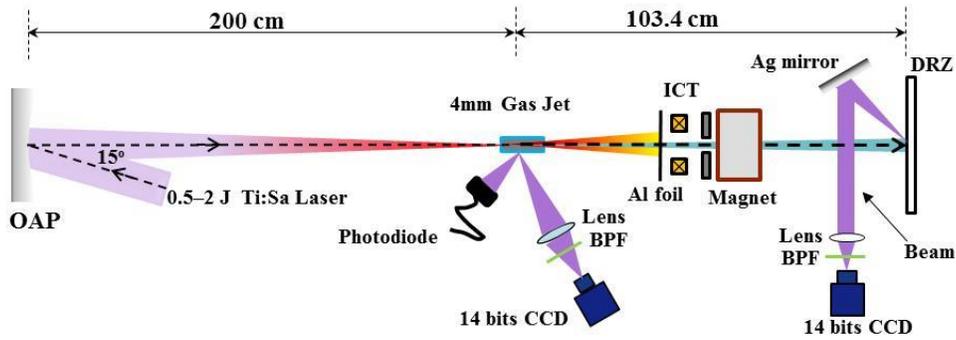

Fig. 1. Schematic diagram of the laser wakefield acceleration experiment at LLP.

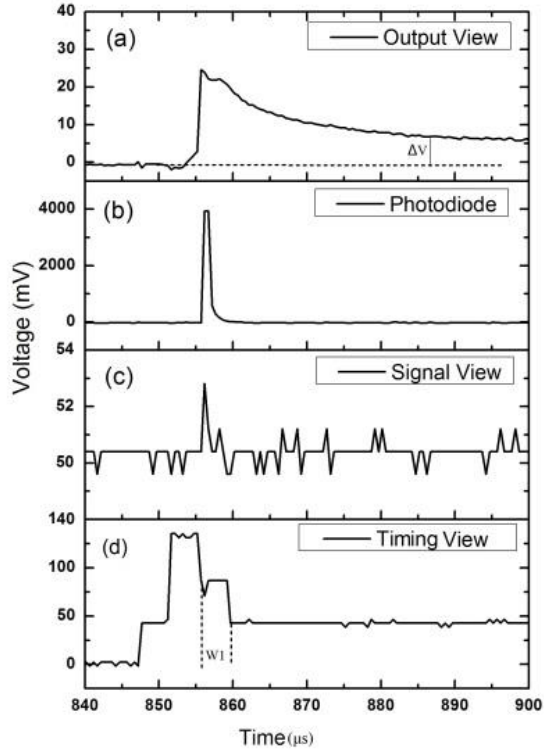

Fig. 2. Wave forms of the beam charge monitoring unit (BCM). (a) Output view of BCM unit, the voltage offset corresponds to the electron beam charge for a given device gain. (b) Directly comes from the fast photodiode showing exact laser-plasma interaction time. (c) Signal view of BCM unit. (d) Timing View of BCM unit, see text for detailed description.

### 3.2. Processing method

Along with measuring the electron beam charge using the ICT, we recorded the beam spatial profile on a DRZ screen using 14 bit CCD camera. Figure 3 shows the electron beam profiles generated from aforementioned gas jets upon interaction with the above-mentioned laser pulses. Each beam image was converted into an ASCII format using PCO CamWare software. If we consider an image as a 2D matrix, each cell may have possible value of 0 to $2^{14}$. This value is practically the intensity of each pixel of the recorded image. A computer code was written to image-process each electron beam profile. This code first finds the exact place of the beam point in the image then makes crop for a certain area of the image including the electron beam point. For those electron beam profiles which contained more than one electron bunch (point) or let say filamentary electron beam (Fig. 8. shows such kind of beams), the code will consider a larger crop area to cover all the recorded electron beam bunches in the same laser shot. To have more accurate image processing, we removed the background noises due to gamma and hard X-rays. Finally, the code calculates the total intensity of the cropped area. From the ICT charge measurement we had electron beam charge for each shot and from our image processing method we found the corresponding total beam intensity. To do the ICT-DRZ cross-calibration we have fitted a linear curve which is shown in Fig. 4. In our experiment and using this curve we were able to find the corresponding electron beam charge for any recorded electron beam profile.

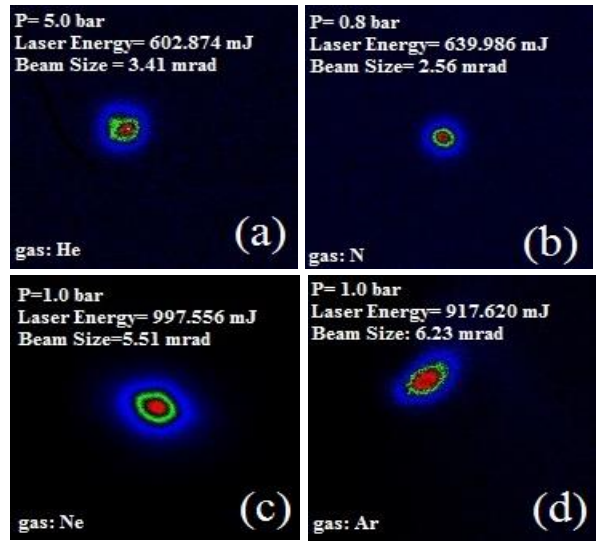

Fig. 3. Electron beam spatial profile from laser-driven plasma of (a) Helium, (b) Nitrogen, (c) Neon, and (d) Argon gas jets.

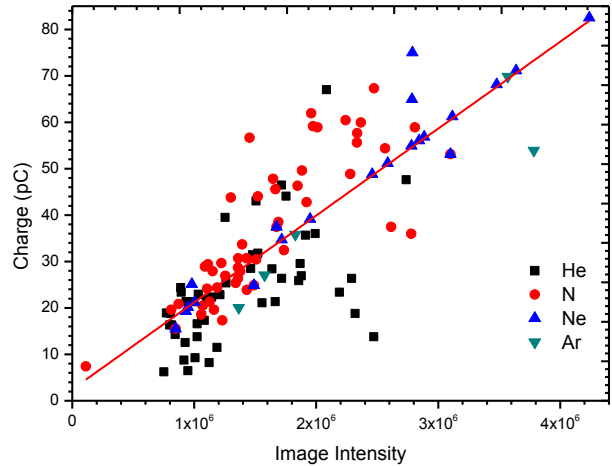

Fig. 4. ICT-DRZ cross-calibration curve for the measurement of the electron beam charge for different gases.

## 4. RESULTS AND DISCUSSION

We obtained the electron beam charge for all the recorded shots on the CCD. We should mention that the laser energy varied (or fluctuated) shot-to-shot, and each of the used gas jets had an optimum pressure range for generating its electron beam. As expected, it is found that the electron beam charge depends on the laser energy and plasma density (or the gas pressure). In the following we will discuss the experimental results for each gas jet individually.

### 4.1. Helium gas jet

Helium is widely used in LWFA experiments as a gas target. In the present experiment we used He gas jet at different stagnation pressures starting from 2 to 25 bars , (corresponds to $n_e$ =3.35×10$^{18}$ cm$^{-3}$ to 4.187×10$^{19}$ cm$^{-3}$ plasma densities where fully ionized helium is guaranteed.[34]) For each pressure step several laser shots were done. He electron beam charge dependence on the laser energy in the range 550 mJ to 1.15 J is shown in Fig. 5. Electron beams were mostly observed at the pressure range 3 to 12 bars (5.02×10$^{18}$ cm$^{-3}$ – 2.01×10$^{19}$ cm$^{-3}$ plasma densities). The highest quality beams (well-collimated of 3.41 mrad) were observed only in pressure of 4.6 to 5.5 bars (Fig. 3a.). In this case, the measured beam charge was ~150 pC or lower. However, a higher electron beam yield (200 pC – 700 pC) was measured at higher laser energy in the range of 900 mJ –1.1 J. We measured the electron beam energy spectrum using the dipole magnet; at the plasma density of 5×10$^{18}$ cm$^{-3}$, the measured energy spectrum is shown in Fig. 6a; up to 300 MeV quasimonoenergtic electron beams are observed from the He gas jet. It was often observed that at a higher gas pressure (6.1 bars corresponding to 10.21×10$^{18}$ cm$^{-3}$ plasma density), a filamentary or spatially-broken (low-quality) electron beams were generated (Fig. 8a.). It was observed, given the relatively-large laser spot size in our experiment, that there is a trend for a *high-quality* (low-quality) and *stable* (unstable) electron beam generation at *low* (higher) plasma density. In addition, the electron self-injection mechanism due to the laser self-focusing is expected to be the main mechanism in this case, as confirmed by our 2D-PIC simulations[36, 37]. Fig.7a. shows the electron beam charge dependence on the plasma density for the He gas jet case. The electron beam charge increases with the density in the range of 5×10$^{18}$ cm$^{-3}$ –1.2×10$^{19}$ cm$^{-3}$, however we have not observed electron beams on the DRZ or ICT at much higher plasma densities. There might be some reasons for this: The first expectation is a very-large electron beam pointing angle (~ 50 mrad at $n_e$ = 2×10$^{19}$ cm$^{-3}$) at high plasma densities[36]. Such a huge pointing angle is almost equal to the solid angle subtended by the DRZ, so at higher densities where the pointing angle is expected to be larger the electron beam might pointed outside the DRZ and thus undetectable. The second expectation might be the short pump-depletion length at very high plasma densities. For instance, the calculated pump depletion length $L_{dp}$ using equation (1)[39] at the plasma density of 1.2×10$^{19}$ cm$^{-3}$ was 0.64 mm and 0.3mm for 2.0×10$^{19}$ cm$^{-3}$. Therefore, we predict that the trapping and acceleration processes were not properly established at those densities and higher. In Eq. (1), $\lambda_p$ is plasma wavelength, $\lambda_L$ is laser wavelength and $a_0$ is normalized laser vector potential.

$$L_{pd} = \frac{2\lambda_p^3}{a_0^2 \lambda_L^2} \quad (1)$$

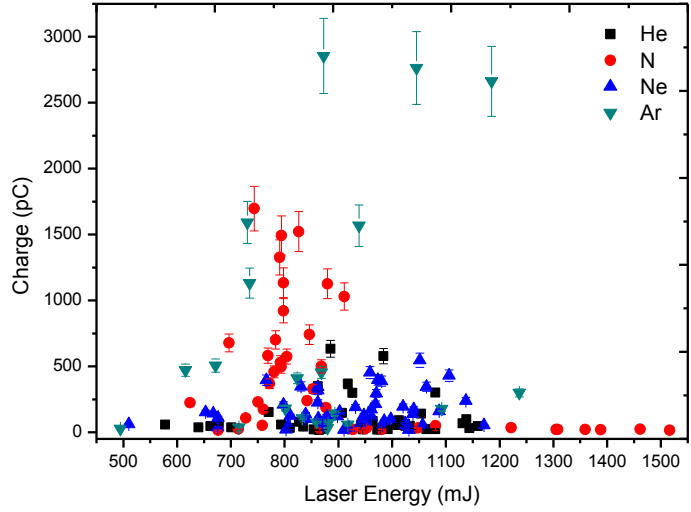

Fig. 5. Electron beam charge dependence on laser energy for He, N, Ne and Ar gases.

### 4.2. Nitrogen gas jet

Nitrogen was used in a few LWFA experiments in the past; the most recent ones were reported in Ref. 37 and Ref. 40. We used N gas jet at the stagnation pressure range 0.5 – 6 bars and the laser energy varied from 600 mJ to 1.6 J. Typical nitrogen's electron beam spatial profile had a divergence of 2.53 mrad which was the smallest in this experiment, as is shown in Fig. 3b. The measured electron beam energy spectrum in this case is shown in Fig. 6b; where a strong monoenergtic peak at 130 MeV is clearly detected. The nitrogen's electron beam charge dependence on the laser energy is shown in Fig. 5, where the charge increases with the laser energy then drops again at high laser energies. The electron beam was mostly observed at pressures between 0.5 to 4 bars ($n_e$ = 0.67×10$^{18}$ cm$^{-3}$ –5.42×10$^{18}$ cm$^{-3}$; assuming the ionization of 5 electrons from each nitrogen atom.[34]) Generally, the measured electron beam charge from nitrogen was high compared with other gases, except for a few laser shots. Electron beam yields as high as 1 nC ~ 2.3nC were measured at the plasma density range of 1.08×10$^{18}$ cm$^{-3}$ –1.49×10$^{18}$ cm$^{-3}$ at a laser energy of ~ 800 mJ. It is rather surprising to observe that at higher laser energy of more than 1 J, the electron beam charge was only ~50 pC or less, yet we have no explanation of this phenomenon. In some shots a high-charge electron beam (2.3 nC) of a large divergence (16 mrad) was observed (Fig. 8b.). We believe that such high beam charge and large divergence angle were due to the formation of nitrogen clusters in the gas jet. To confirm this, we recall the condition of cluster formation in supersonic nozzles: it can be described by an empirical scaling parameter $\Gamma^*$ which is related to what is called "Hagena parameter, k".[41,42]

$$\Gamma^* = k \frac{(d/\tan\alpha)^{0.85}}{T_0^{2.29}} P_0 \qquad (2)$$

Where *d* is the nozzle diameter (at throat) in µm (*d* = 120 µm in our nozzle), α is the expansion half angle (for our supersonic gas jet α = 4°), $P_0$ is backing plenum pressure, and $T_0$ is pre-expansion temperature in Kelvin ($T_0$ =293 K). The Hagena parameter for nitrogen in our case, k = 528.[42] Clustering generally begins for $\Gamma^* > 1000$.[41,42] In the present experiment for the nitrogen case at $P_0$ >1.23 bar ($n_e$ = 1.67×10$^{18}$ cm$^{-3}$) we found $\Gamma^* > 1005$ which means that there is a high chance to form nitrogen clusters in our gas jet.

Figure 7b shows the charge dependence on the plasma density for the nitrogen gas jet. The beam charge has increased with the plasma density in range of 6.7×10$^{17}$ cm$^{-3}$ to 1.49×10$^{18}$ cm$^{-3}$ then it dropped at higher plasma densities. We used the relation (1) to calculate the laser pump depletion length in this case. For the plasma density of 2×10$^{18}$ cm$^{-3}$ (corresponding to gas pressure 1.5 bars) the $L_{dp}$ was 9.4 mm which is much longer than our gas jet length. It means that for the nitrogen case for up to the plasma density of $3.5 \times 10^{18}$ cm$^{-3}$ at which the pump depletion was 4mm, the electron beam charge should increase with the plasma density. However, as shown in Fig.7b, the electron beam charge drops at plasma density higher than $1.5 \times 10^{18}$ cm$^{-3}$, at which it is highly likely that nitrogen clusters were formed. It seems that the relation (1) for pump depletion is not valid in case of clustered gas jet. The pump depletion length seemed to be extremely short in the clustered gas case; therefore the trapping and acceleration processes could not be established for a long distance.

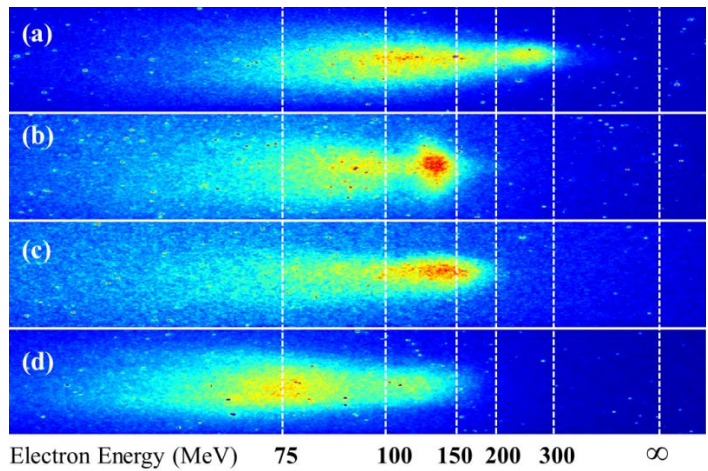

Fig. 6. Electron beams energy from (a) Helium gas, plasma density 5.02×10$^{18}$ cm$^{-3}$, (b) Nitrogen gas, plasma density 0.27×10$^{18}$ cm$^{-3}$ (c) Neon gas, plasma density 1.34×10$^{18}$ cm$^{-3}$ (d) and Argon gas, plasma density 1.76×10$^{18}$ cm$^{-3}$.

### 4.3. Neon gas jet

Neon gas jet was the third target used in our experiment; in this case the working gas pressure range was 0.3 –13 bars. It's the first experience with LWFA in neon; therefore it is rather interesting to find out what electron beams could be generated in this case. Ne electron beam charge dependence on the laser energy is shown in Fig. 5, the electron beam charge reached ~ 300 pC in this case. In this experiment, we found that the generation of electron beams from the Ne gas difficult and sensitive to the laser energy if compared with other gas targets. At laser energies below 700 mJ, no electrons were observed. In most laser shots where high-quality collimated neon electron beams were observed (Fig. 3c.), the laser energy was in the range of 800 mJ –1.1 J and the gas pressure range was 1– 4 bars, (corresponding to $n_e$ = 6.7×10$^{18}$ cm$^{-3}$ to 2.68×10$^{19}$ cm$^{-3}$; assuming 6 electrons were released from each Ne atom upon interaction with the above-mentioned laser pulses. [34, 43]) The maximum electron beam energy measured in this case was about 200 MeV with quasimonoenergtic structures around 170 MeV, as shown in Fig. (6c.). In few shots with lower laser energy (<800 mJ) and at the gas pressure 10 to 12 bars unstable and filamentary electron beams were observed (Fig. 8c). The electron yield dependence on the plasma density for the neon gas jet is shown in Fig.7c. The beam charge increases with the plasma density then decreases at higher densities (5×10$^{19}$ cm$^{-3}$). Using the dephasing equation (1) for the neon gas jet at a plasma density more than 5×10$^{19}$ cm$^{-3}$ shows the pump depletion length was shorter than 75.7 μm. This means the depletion length will be too short at those high plasma densities to sufficiently trap and accelerate the electrons in this case.

### 4.4. Argon gas

In some previous LWFA experiments, argon mixed with other gas were used to realize the ionization injection[44], it was also used alone as a target.[45,46] In the present experiment the argon gas jet stagnation pressure was in range 0.5 to 10 bars. Collimated electron beams (Fig. 3d.) were observed at the laser energy of 700 mJ to 1 J at the pressure range of 0.8 –1.2 bars (corresponding to $n_e$ = 4.64×10$^{18}$ cm$^{-3}$ –7.03×10$^{18}$ cm$^{-3}$, assuming 8 electrons were released from each Ar atom.[34,43]) The electron beam energy from the Ar gas jet was the lowest among all, however it shows quasimonoenergtic structures around 75 MeV and the maximum energy reached 150 MeV, as shown in Fig. 6d. The argon electron beam charge versus the laser energy is shown in Fig. 5. For most laser shots the electron beam charge was <500 pC. Higher beam charge up to 2.7 nC was measured in some shots when the filamentary of electrons beam were observed (Fig. 8d.). Using relation (2) it is found that $G^*$ >1275 for the Ar gas jet given k = 1650,[42] $T_0$ = 293K and $P_0$ > 0.5 bar which satisfies the condition of cluster formation. A supporting evidence for the cluster formation in this case was the observed conical laser-plasma channels (Fig. 9a). The channel started narrow at the edge of the nozzle, then gradually became wider and wider. Compared with a usual laser-plasma channel (for example in the helium case (Fig. 9b)), the laser-plasma channel in the Ar case have shown a very different shape which might be a characteristic of cluster formation or some kind of Ar gas-cluster mixture. In this case, typically the beam charge is expected to be very high (Fig. 4, for the Ar and N) as observed in our previous work.[47] The high-charge generated in this case was attributed to a combined LWFA and direct laser acceleration (DLA) mechanisms.[47] Furthermore, the conical shape of the argon laser-plasma channel qualitatively resembles a splash-like channel observed experimentally,[46] in laser-argon interaction (at nonrelativistic laser intensities) in an identical gas jet. The utilization of this kind of laser-plasma channel in the argon for laser wakefield acceleration has been studied by 2D PIC-simulations in the same paper,[46] it was found that such kind of channels are potential candidates for well-collimated, high-charge, and small energy-spread electron beams, all those features are observed in our experiment employing argon gas jet, as we discussed earlier in details.

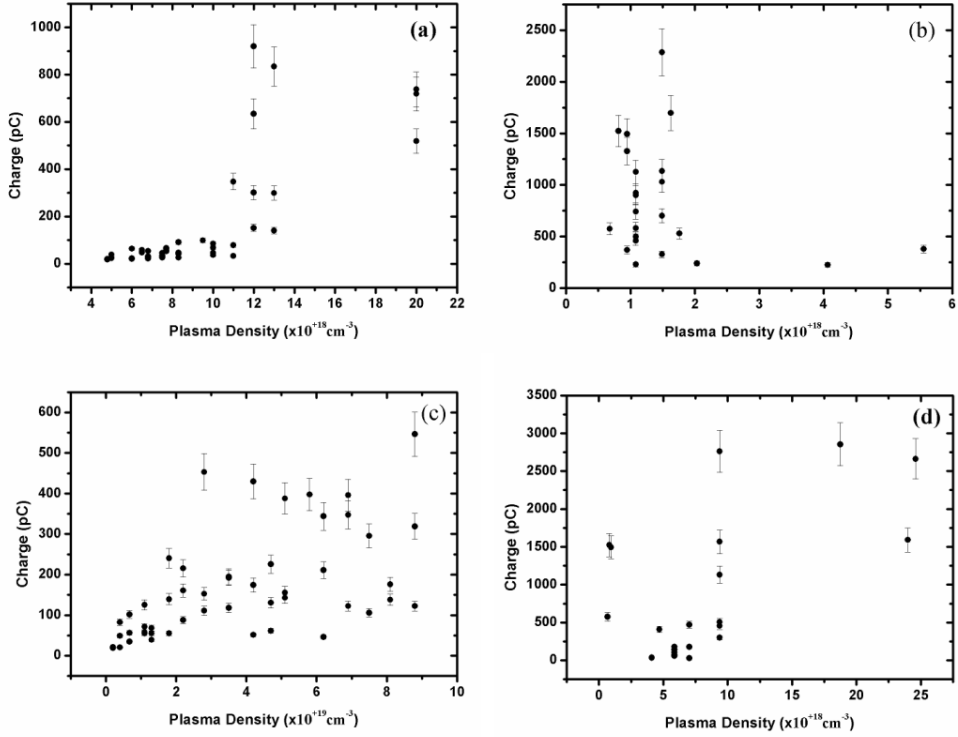

Fig. 7. Achieved electron beam charge and its dependence on the plasma density for (a) He, (b) N, (c) Ne and (d) Ar gas jets

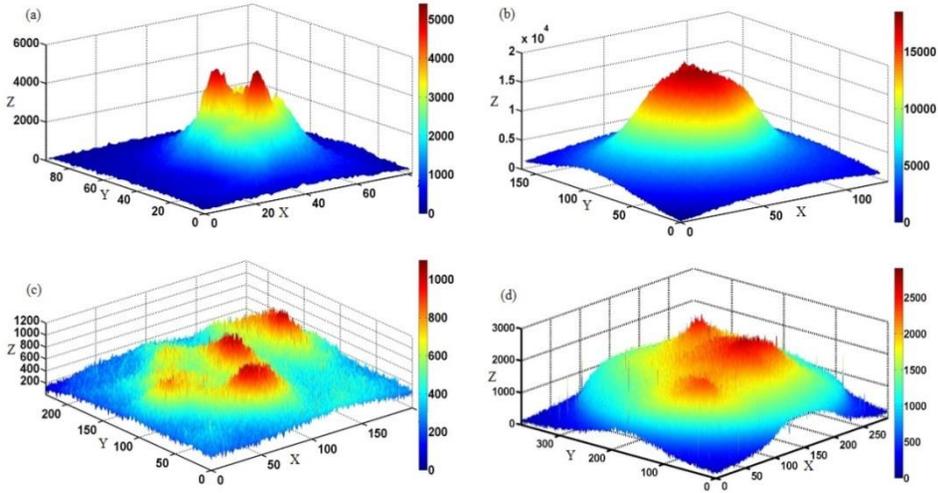

Fig. 8. 3D image of electron beam profile. Z-axis shows beam intensity. (a) filamentary He electron beam in $10.21 \times 10^{18}$ cm$^{-3}$ plasma density at laser energy 906.9 mJ, (b) Nitrogen gas with electron beam charge of 2.28 nC at $1.49 \times 10^{18}$ cm$^{-3}$ plasma density, (c) Neon gas with $6.76 \times 10^{18}$ cm$^{-3}$ plasma density and laser energy of 766 mJ (d) Argon gas, electron beam charge of 2.7 nC at laser energy 1.04 J and $9.37 \times 10^{18}$ cm$^{-3}$ plasma density.

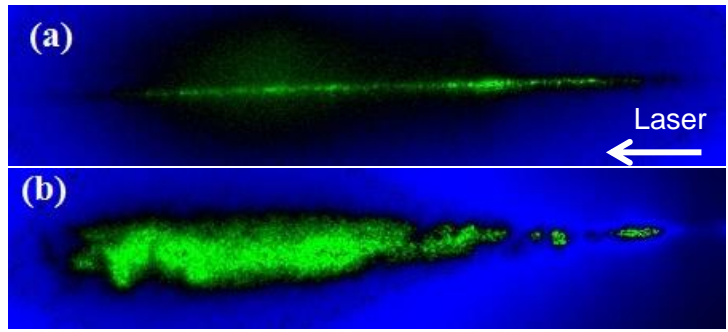

Fig. 9. Plasma channel (a) Well-collimated narrow laser-plasma channel in the case of He gas jet. (b) A conic shaped laser-plasma channel in the case of Ar gas jet.

## 5. CONCLUSIONS

By investigating electron beams from laser driven He, N, Ne and Ar gas jets individually we could achieve collimated beams with multi-hundred MeV energies and high yields. As shown in Fig. (5), and within the available laser beam energy range of 0.5–1.5 J, the highest beam charge of~ 3nC was achieved from the Ar gas jet. Nitrogen gas jet produced as high as 1.7 nC electron beams. Helium and neon gas jets produced as high as ~ 500 pC electron beams. However, the beams with the highest charge were mostly filamentary with large divergence angles. The charge of the well-collimated electron beams were modest 10-100 pC for all the used gases. The working plasma density range for the four gases were different; for He and Ne gases at high densities the beam charge dropped potentially due to the very strong pump depletion. As for the Ar and N gases, cluster formation was likely occurred, especially in the Ar gas case where we observed a very different shape (conic or splash-like) of the plasma channel. In that case (Ar and N clusters) the laser acceleration could generate electron beams with a very high charge due to a combination of LWFA and DLA (direct laser acceleration) mechanisms.[47]

### Acknowledgment

This work was supported by the 973 National Basic Research Program of China (Grant No. 2013CBA01504), and the Natural Science Foundation of China NSFC (Grants: 11121504, 11334013, 11175119, and 11374209).

### References

1. T. Tajima and J. M. Dawson, "Laser electron acceleration," Phys. Rev. Lett. **43**, 267–270 (1979); T. Tajima, "Laser wakefield: Bring accelerators down to size," Nature Photon. **2**, 526–527 (2008).
2. G. A. Mourou, T. Tajima, and S. V. Bulanov, "Optics in the relativistic regime," Rev. Mod. Phys. **78**, 309–371 (2006).
3. N. A. M. Hafz, T. M. Jeong, I. W. Choi, S. K. Lee, K. H. Pae, V. V. Kulagin, J. H. Sung, T. J. Yu, K.-H. Hong, T. Hosokai, J. R. Cary, D.-K. Ko, and J. Lee, "Stable generation of GeV-calss electron beams from self-guided laser-plasma channels," Nature Photon. **2**, 571–577 (2008).
4. C. G. R. Geddes, C. Toth, J. van Tilborg, E. Esarey, C. B. Schroeder, D. Bruhwiler, C. Nieter, J. Cary, and W. P. Leemans, "High-quality electron beams from a laser wakefield accelerator using plasma-channel guiding," Nature **431**, 538–541 (2004).
5. S. Mangles, C. D. Murphy, Z. Najmudin, A. G. R. Thomas, J. L. Collier, A. E. Dangor, E. J. Divall, P. S. Foster, J. G. Gallacher, C. J. Hooker, D. A. Jaroszynski, A. J. Langley, W. B. Mori, P. A. Norreys, F. S. Tsung, R. Viskup, B. R. Walton, and K. Krushelnick, "Monoenergetic beams of relativistic electrons from intense laser plasma interactions," Nature **431**, 535–538 (2004).
6. J. Faure, Y. Glinec, A. Pukhov, S. Kiselev, S. Gordienko, E. Lefebvre, J.-P. Rousseau, F. Burgy, and V. Malka, "A laser-plasma accelerator producing monoenergetic electron beams," Nature **431**, 541–544 (2004).
7. K. Nakamura, B. Nagler, C. Toth, C. G. R. Geddes, C. B. Schroeder, E. Esarey, W. P. Leemans, A. J. Gonsalves, and S. M. Hooker, "GeV electron beams from a centimeter-scale channel guided laser wakefield accelerator," Phys. Plasmas **14**, 056708 (2007).

8. A. J. Gonsalves,ha T. P. Rowlands-Rees, B. H. P. Broks, J. J. A. M. van der Mullen, and S. M. Hooker, "Transverse interferometry of a hydrogen-filled capillary discharge waveguide," Phys. Rev. Lett. **98**, 025002 (2007).
9. W. P. Leemans, B. Nagler, A. J. Gonsalves, Cs. Tóth, K. Nakamura, C. G. R. Geddes, E. Esarey, C. B. Schroeder, and S. M. Hooker, "GeV electron beams from a centimeter-scale accelerator," Nature Phys. **2**, 696–699 (2006).
10. X. Wang, R. Zgadzaj, N. Fazel, Z. Li, S. A. Yi, X. Zhang, W. Henderson, Y. Y. Chang, R. Korzekwa, H. E. Tsai, C. H. Pai, H. Quevedo, G. Dyer , E. Gaul, M. Martinez, A. C. Bernstein, T. Borger, M. Spinks, M. Donovan, V. Khudik, G. Shvets, T. Ditmire, and M. C. Downer, "Quasi-monoenergetic laser-plasma acceleration of electrons to 2 GeV," Nature Comm. **4**, 1988 (2013).
11. H. Taek, K. Ki, H. Pae, S. K. Lee, H. J. Cha, T. M. Jeong, I. J. Kim, T. J. Yu, J. Lee, and J. H. Sung, "Enhancement of electron energy to the multi-GeV regime by a dual-stage laser-wakefield accelerator pumped by petawatt laser pulses," Phys. Rev. Lett. **111**, 165002 (2013).
12. P. A. Walker, N. Bourgeois, W. Rittershofer, J. Cowley, N. Kajumba, A R Maier, J. Wenz, C. M. Werle, S. Karsch, F. Gruner, D. R. Symes, P. P. Rajeev, S. J. Hawkes, O. Chekhlov, C. J. Hooker, B. Parry, Y. Tang, and S. M. Hooker, "Investigation of GeV-scale electron acceleration in a gas-filled capillary discharge waveguide," New J. Phys. **15**, 045024 (2013).
13. K. Nakajima, H. Lu, X. Zhao, B. Shen, R. Li, and Z. Xu, "100-GeV large scale laser plasma electron acceleration by a multi-PW laser," Chin. Opt. Lett. **11**, 013501 (2013).
14. M. Z. Mo, A. Ali, S. Fourmaux, P. Lassonde, J. C. Kieffer, and R. Fedosejevs, "Generation of 500 MeV–1 GeV energy electrons from laser wakefield acceleration via ionization induced injection using $CO_2$ mixed in He," App. Phys. Lett. **102**, 134102 (2013).
15. S. Corde, K. Ta Phuoc, G. Lambert, R. Fitour, V. Malka, and A. Rousse, A. Beck, and E. Lefebvre, "Femtosecond x rays from laser-plasma accelerators," Rev. Mod. Phys. **85**, 1–46 (2013).
16. A. Ben-Ismail, J. Faure, and V. Malka, "Optimization of gamma-ray beams produced by a laser-plasma accelerator," Nuc. Ins. Meth. Phys. Res. A **629**, 382–386 (2011).
17. A. Ben-Ismail, O. Lundh, C. Rechatin, J. K. Lim, J. Faure, S. Corde, and V. Malka, "Compact and high-quality gamma-ray source applied to 10 µm-range resolution radiography," App. Phys. Lett. **98**, 264101 (2011).
18. Y. Wu, Z. Zhao, B. Zhu, K. Dong, X. Wen, Y. He, Y. Gu, and B. Zhang, "Laser wakefield electron acceleration for γ-ray radiography application," Chin. Opt. Lett. **10**, 063501 (2012).
19. N. D. Powers, I. Ghebregziabher, G. Golovin, C. Liu, S. Chen, S. Banerjee, J. Zhang, and D. P. Umstadter, "Quasi-monoenergetic and tunable X-rays from a laser-driven Compton light source," Nature Photon. **8**, 28–31 (2014).
20. S. Chen, N. D. Powers, I. Ghebregziabher, C. M. Maharjan, C. Liu, G. Golovin, S. Banerjee, J. Zhang, N. Cunningham, A. Moorti, S. Clarke, S. Pozzi, and D. P. Umstadter, "MeV-energy X rays from inverse Compton scattering with laser-wakefield accelerated electrons," Phys. Rev. Lett. **110**, 155003 (2013).
21. D. Umstadter, S.-Y. Chen, A. Maksimchuk, G. Mourou, and R. Wagner, "Nonlinear optics in relativistic plasmas and laser wake field acceleration of electrons," Science **273**, 472–475 (1996).
22. C. Gahn, G. D. Tsakiris, K. J. Witte, P. Thirolf, and D. Habs, "A novel 45-channel electron spectrometer for high intensity laser-plasma interaction studies," Rev. Sci. Instrum. **71**, 1642–1645 (2000).
23. M. Galimberti, A. Giulietti, D. Giulietti, and L. A. Gizzi, "SHEEB: A spatial high energy electron beam analyzer," Rev. Sci. Instrum. **76**, 053303 (2005).
24. K. Nakamura, A. J. Gonsalves, C. Lin, A. Smith, D. Rodgers, R. Donahue, W. Byrne, and W. P. Leemans, "Electron beam charge diagnostics for laser plasma accelerators," Phys. Rev. ST Accel. Beams **14**, 062801 (2011).
25. K. Zeil, S. D. Kraft, A. Jochmann, F. Kroll, W. Jahr, U. Schramm, L. Karsch, J. Pawelke, B. Hidding, and G. Pretzler, "Absolute response of Fuji imaging plate detectors to picosecond-electron bunches," Rev. Sci. Instrum. **81**, 013307 (2010)
26. N. Hafz, M. S. Hur, G. H. Kim, C. Kim, I. S. Ko, and H. Suk, "Quasimonoenergetic electron beam generation by using a pinholelike collimator in a self-modulated laser wakefield acceleration," Phys. Rev. E **73**, 016405 (2006).
27. Y. Glinec, J. Faure, A. Guemnie-Tafo, and V. Malka, "Absolute calibration for a broad range single shot electron spectrometer," Rev. Sci. Instrum. **77**, 103301 (2006).
28. Y. C. Wu, B. Zhu, K. G. Dong, Y. H. Yan, and Y. Q. Gu, "Note: Absolute calibration of two DRZ phosphor screens using ultrashort electron bunch," Rev. Sci. Instrum. **83**, 026101 (2012).
29. T. Fuchs, H. Szymanowski, U. Oelfke, Y. Glinec, C. Rechatin, J. Faure, and V. Malka, "Treatment planning for laser-accelerated very-high energy electrons" Phys. Med. Biol. 54, 3315 (2009).
30. T. Hosokai, K. Kinoshita, T. Watanabe, K. Yoshii, T. Ueda, A. Zhidkov, M. Uesaka, K. Nakajima, M. Kando, and H. Kotaki, "Supersonic gas jet target for generation of relativistic electrons with 12TW-50fs laser" in Proceedings of 8th European Particle Accelerator Conference (EPAC), Paris, France, 3–7 June (European Physical Society, Geneva, 2002), pp. 981-983.
31. W. Yan, L. Chen, D. Li, L. Zhang, N. A. M. Hafz, J. Dunn, Y. Ma, K. Huang, Luning Su, M. Chen, Z. Sheng, J. Zhang, " concurrence of monoenergetic electron beams and bright X-rays from an evolving laser plasma bubble" Proc. Natl. Acad. Sci. USA 111, 5825 (2014).
32. N. M. Hafz, I. W. Choi, J. H. Sung, H. T. Kim, K.-H. Hong, T. M. Jeong, and T. J. Yu, "Dependence of the electron beam parameters on the stability of laser propagation in alaser wakefield accelerator" App. Phys. Lett. **90,** 151501 (2007).

33. R. Fedosejevs, X. F. Wang, and G. D. Tsakiris, "Onset of relativistic self-focusing in high density gas jet targets" Phys. Rev. E 56, 4615 (1997).
34. P. Gibbon, *Short Pulse Laser Interactions with Matter* (Imperial College Press, London, 2005) p. 20-23
35. H. J. Cha, I. W. Choi, H. T. Kim, I. J. Kim, K. H. Nam, T. M. Jeong, and J. Lee, "Absolute energy calibration for relativistic electron beams with pointing instability from a laser-plasma accelerator," Rev. Sci. Instrum. **83**, 063301 (2012).
36. S. Li, N. A. M. Hafz, M. Mirzaie, X. Ge, T. Sokollik, M. Chen, Z. Sheng, and J. Zhang, "Stable laser-plasma accelerators at low densities," J. Appl. Phys. **116**, 043109 (2014).
37. M. Tao, N. A. M. Hafz, S. Li, M. Mirzaie, A. M. M. Elsied, X. Ge, F. Liu, T. Sokollik, L. Chen, Z. Sheng, and J. Zhang, "Quasimonoenergetic collimated electron beams from a laser wakefield acceleration in low density pure nitrogen," Phys. Plasmas. **21**, 073102 (2014).
38. S. Li, N. A. M. Hafz, M. Mirzaie, A. M. M. Elsied, X. Ge, F. Liu, T. Sokollik, M. Tao, L. Chen, M. Chen, Z. Sheng, and J. Zhang, "Generation of electron beams from a laser wakefield acceleration in pure neon gas," Phys. Plasmas **21**, (2014).
39. E. Esarey, C. B. Schroeder, and W. P. Leemans, "Physics of laser-driven plasma-based electron accelerators," Rev. Mod. Phys. **81**, 1229–1285 (2009).
40. M. Z. Mo, A. Ali, S. Fourmaux, P. Lassonde, J. C. Kieffer, and R. Fedosejevs, "Quasimonoenergetic electron beams from laser wakefield acceleration in pure nitrogen," App. Phys. Lett. **100**, 074101 (2012).
41. O. F. Hagena, "Silver cluster from nozzle expansions," Z. Phys. D – Atoms, Molecules and Clusters **17**, 157–158 (1990).
42. R. A. Smith, T. Ditmire, and J. W. G. Tisch, "Characterization of a cryogenically cooled high-pressure gas jet for laser/cluster interaction experiments," Rev. Sci. Instrum. **69**, 3798–3804 (1998).
43. R. Fedosejevs, X. F. Wang, and G. D. Tsakiris, "Onset of relativistic self-focusing in high density gas jet targets," Phys. Rev. E **56**, 4615–4639 (1997).
44. Y.-C. Ho, T.-S. Hung, C.-P. Yen, S.-Y. Chen, H.-H. Chu, J.-Y. Lin, J. Wang, and M.-C. Chou, "Enhancement of injection and acceleration of electrons in a laser wakefield accelerator by using an argon-doped hydrogen gas jet and optically preformed plasma waveguide," Phys. Plasmas **18**, 063102 (2011).
45. M. Mori, K. Kondo, Y. Mizuta, M. Kando, H. Kotaki, M. Nishiuchi, M. Kado, A. S. Pirozhkov, K. Ogura, H. Sugiyama, S. V. Bulanov, K. A. Tanaka, H. Nishimura, and H. Daido, "Generation of stable and low-divergence 10-MeV quasimonoenergetic electron bunch using argon gas jet," Phys. Rev. ST Accel. Beams **12**, 082801 (2009).
46. Y. Mizuta, T. Hosokai, S. Masuda, A. Zhidkov, K. Makito, N. Nakanii, S. Kajino, A. Nishida, M. Kando, M. Mori, H. Kotaki, Y. Hayashi, S. V. Bulanov, and R. Kodama, "Splash plasma channels produced by picosecond laser pulses in argon gas for laser wakefield acceleration," Phys. Rev. ST Accel. Beams **15**, 121301 (2012).
47. L. M. Chen, W. C. Yan, D. Z. Li, Z. D. Hu, L. Zhang, W. M. Wang, N. Hafz, J. Y. Huang, Y. Ma, J. R. Zhao, J. L. Ma, Y. T. Li, X. Lu, Z. M. Sheng, Z. Y. Wei, J. Gao, and J. Zhang, "Bright betatron X-ray radiation from a laser-driven-clustering gas target," Sci. Rep. **3**, 1912 (2013).